\documentclass[useAMS,usenatbib]{mn2e}

\usepackage{graphicx}
\usepackage{ulem}

\def\be{\begin{equation}}
\def\ee{\end{equation}}
\def\ba{\begin{eqnarray}}
\def\ea{\end{eqnarray}}

\def\ltsima{$\; \buildrel < \over \sim \;$}
\def\simlt{\lower.5ex\hbox{\ltsima}}
\def\gtsima{$\; \buildrel > \over \sim \;$}
\def\simgt{\lower.5ex\hbox{\gtsima}}
\def\etal{{et al.\ }}

\usepackage[dvipsnames]{color}

\title[Cooling rate for a collisionally cooled gas]
{Non-equilibrium cooling rate for a collisionally cooled metal-enriched gas}
\author[E. O. Vasiliev]
       {Evgenii O. Vasiliev$^{1}$\thanks{E-mail:eugstar@mail.ru}\\
$^1$Institute of Physics, Department of Physics, Southern Federal University, Stachki Ave. 194,        
Rostov-on-Don, 344090 Russia\\
}
\begin{document}
\date{Accepted 3004 December 15.
      Received 2004 December 14;
      in original form 2004 December 31}
\pagerange{\pageref{firstpage}--\pageref{lastpage}}
\pubyear{3004}
\maketitle

\label{firstpage}

\begin{abstract}
We present self-consistent calculations of non-equilibrium (time-dependent) cooling rates for a dust-free 
collisionally controlled gas in wide temperature ($10~K\le T\le 10^8~K$) and metallicity ($10^{-4}~Z_\odot 
\le Z \le 2~Z_\odot$) ranges. We confirm that molecular hydrogen dominates cooling at $10^2 \simlt T\simlt 
10^4$~K and $Z\simlt 10^{-3}~Z_\odot$. We find that the contribution from H$_2$ into cooling rate around 
$T\sim (4-5)\times 10^3$~K stimulates thermal instability in the metallicity range $Z\simlt 10^{-2}~Z_\odot$. 
Isobaric cooling rates are generally lower than isochoric ones, because the associated increase of gas density 
leads to both more efficient hydrogen recombination and equilibration of the fine-structure level populations. 
Isochoric cooling keeps the ionization fraction remains quite high at $T\simlt10^4$~K: up to $\sim0.01$ at 
$T\simeq 10^3$~K and $Z\simlt 0.1~Z_\odot$, and even higher at higher metallicity, contrary to isobaric 
cooling where it at least an order of magnitude lower. Despite this increase in ionization fraction the 
gas-phase formation rate of molecular hydrogen (via H$^-$) lowers with metallicity, because higher metallicity
shorttens the evolution time. We implement our self-consistent cooling rates into the multi-dimensional parallel 
code ZEUS-MP in order to simulate evolution of a supernova remnant, and compare it with an analogous model with 
tabulated cooling rates published in previous works. We find significant differences between the two descriptions, 
which may appear, e.g., in mixing of the ejected metals in the circumstellar medium. 
\end{abstract}



\section{Introduction}

\noindent

A choice between equilibrium and non-equilibrium (time-dependent) cooling rates in collisionally controlled 
gas strongly depends on the ratio between dynamical and chemical/cooling times. Calculations of the cooling 
rates of astrophysical plasma in the collisional ionization equilibrium (CIE) for $T\simgt 10^4$~K were performed 
by many authors \citep{house,cox,raymond,shull82,gaetz,bohringer,sd93,landi,benjamin,bryans,schure09}.
However, calculations of the time-de\-pendent ionization of metals and associated radiative cooling rates showed 
significant deviations from the CIE states and rates \citep{kafatos,shapiro,edgar,schmutzler,sd93,gs07,avillez2,v11}.

A more complex task is a calculation of cooling rate below $10^4$~K, because we need to take into account 
many processes between atomic and molecular species \citep[e.g.,][]{dalgarno72,hmk79,hmk89}. Apart from these 
difficulties a level of the calculation complexity of the cooling rate depends on one's purposes and typical
timescales of physical problem considered. Under some conditions chemical network at $T<10^4$~K 
can be constrained significantly. For example, \citet{abel97} and \citet{galli98} proposed the reduced chemical 
network for the primordial gas, \citet{glover07} considered the chemistry of the metal-enriched interstellar 
medium. Sometimes, only cooling rates (without solving chemical kinetics equations) are used to study the 
interstellar medium, for instance, the rates calculated by \citet{dalgarno72} are applied widely.

Usually the cooling rates are provided separately in two temperature ranges, the first is $T=10^4-10^8$~K \citep[e.g.,][]{raymond,sd93,bryans,gs07} and the second is $T\simlt 10^4$~K \citep[e.g.,][]{dalgarno72,hmk89}. 
Very rarely the rates are presented in both ranges \citep[e.g.,][]{sn97}. This is explained by difference 
{ in physical problems considered by authors. But very frequently one needs cooling rates in both ranges. 
So that usually high- and low-temperature cooling rates are combined from different tables, where the rates were 
computed using different assumptions.} Thus, it is necessary to conduct a self-consistent calculation of cooling 
rate within both high- and low-temperature ranges.

In this paper we present self-consistent calculations of non-equilibrium cooling rates of a collisionally cooled
metal-enriched dust-free gas in wide temperature range, $10~K\le T\le 10^8~K$, and we study dependence of the 
rates on metallicity and density. A possible influence from dust grains and external ionizing radiation field will 
be considered elsewhere. The paper is organized as follows. In Section 2 we briefly describe the details of the model.
In Sections 3 and 4 we present our results. { In Section 5 we study the supernova shell evolution using newly
calculated cooling rates.} In Section 6 we summarize our results.


\section{Model description}

\noindent

The chemical and thermal evolution of a gas parcel can be divided into a high-temperature 
($T> 2\times 10^4$~K) and a low-temperature ($T\le 2\times 10^4$~K) ranges. {  Such division
can be explained by the transition to neutral gas and the formation of molecules in the latter range.}
The full description of the method and the references to the atomic data for the former temperature range 
can be found in \citet{v11}. Briefly, in our calculations we consider all ionization states of the elements 
H, He, C, N, O, Ne, Mg, Si and Fe. We take into account the following major processes in a collisional gas: 
collisional ionization, radiative and dielectronic recombination as well as charge transfer in 
collisions with hydrogen and helium atoms and ions. 

To study the chemical evolution of a low-temperature ($T\le 2\times 10^4$~K) gas we should take into
account molecules, especially molecular hydrogen, which is expected to be the most significant coolant 
in a gas with metallicity below $10^{-3}Z_\odot$. Hence, the above-listed ionization states of the 
elements is replenished by a standard set of species: H$^-$, H$_2$, H$_2^+$, D, D$^+$, D$^-$, HD, needed to model 
the H$_2$/HD gas-phase kinetics \citep{abel97,galli98}. The corresponding rates are taken from
\citet{galli98,stancil98,maclow86}. The list of the chemical reactions for molecular hydrogen and
deuterium chemistry is presented in Appendix~1.

The system of time-dependent ionization state equations should be complemented by the energy equation,
{ which can be written in the form }
\be
 {du\over dt} = {p\over \rho^2} {d\rho \over dt} - L
 \label{dq}
\ee
{ 
where $u = p/(\gamma-1)\rho$ is the thermal energy, $p$ is the gas pressure, $\rho$ is the gas density,
$\gamma$ is the adiabatic index, and $L$ is the generalized energy loss function.
The equation (\ref{dq}) can be expressed as 
}
\be
 {1 \over \gamma-1}{dp \over dt} - {\gamma \over \gamma-1} {p \over \rho} {d\rho \over dt} = - L\rho
\ee
{
Assuming $\gamma=5/3$ we find that the gas temperature evolves as \citep[e.g., ][]{kafatos}
}
\be
 {dT \over dt} = -{ n_e n_H  \Lambda \over A n k_B} - {T\over n}{dn\over dt}
 \label{tevol}
\ee
where $n$ is the total particle density of the gas, $n_e$ and $n_H$ are the electron and total hydrogen number
densities, $\Lambda(x_i,T,Z)$ is cooling rate, $k_B$ is the Boltzman constant, $A$ is a constant equal to $3/2$ 
for isochoric and $5/2$ for isobaric cooling. The second term in equation (\ref{tevol}) reflects the relative 
change in the number of particles in the unit of volume, such change is very important in gas below $T\simlt 10^4$~K, 
where gas recombines rapidly.

\begin{table}
\caption{Solar elemental abundances}
\center
\begin{tabular}{lc}
\hline
\hline
   Element  &    (X/H)$_\odot$  \\
\hline
{ Helium}  &  $0.081$    \\
Carbon      &  $2.45\times 10^{-4}$    \\
Nitrogen    &  $6.03\times 10^{-5}$    \\
Oxygen      &  $4.57\times 10^{-4}$    \\
Neon        &  $1.95\times 10^{-4}$    \\
Magnesium   &  $3.39\times 10^{-5}$  \\
Silicon     &  $3.24\times 10^{-5}$   \\
Iron        &  $2.82\times 10^{-5}$    \\
\hline
\hline
\end{tabular}%
\label{table1}
\end{table}

The total (except molecular deposit) cooling rate is calculated using the photoionization code {\small CLOUDY} 
\citep[ver. 08.00,][]{cloudy}. The radiative losses by H$_2$ is taken from \citep{galli98} and the HD 
cooling rate is taken from \citep{flower00,lipovka05}. For the solar metallicity we adopt the abundances 
reported by \citet{asplund}, except Ne for which the enhanced abundance is adopted \citep{drake},
{ the solar abundances are listed in Table~\ref{table1}}. In our calculations we assume the helium 
mass fraction $Y_{\rm He} = 0.24$, { which corresponds to [He/H]=0.081. Such value is close to observed 
one \citep{izotov}. In general, it would be difficult at this time to exclude any value of $Y_{\rm He}$ 
inside the range $0.232-0.258$ \citep[e.g.][]{olive}.} We solve a set of 103 coupled equations (102 for 
ions, atoms and molecules, and one for temperature) using a Variable-coefficient Ordinary Differential Equation 
solver\footnote{The numerical code is available from http://www.netlib.org/ode/vode.f} \citep{dvode}. 

In our calculations we don't include the chemical kinetics of carbon monoxide and lithium hydrid, dust and 
photoionization effects. It is well known that CO and H$_2$O molecules significantly affect on the thermal 
state of the interstellar gas \citep[e.g.,][]{hmk79,omukai05,glover07}. The CO and H$_2$O cooling rates 
dominate over the C and O fine-structure cooling in gas with $T\simgt 500$~K \citep[e.g.,][]{glover07}, but 
whether it is possible to produce enough CO and H$_2$O molecules in the gas during its cooling timescale. 
In the next section we show that the CO/H$_2$O formation timescale is greater than the cooling time, so that 
such molecules cannot form in significant amounts within timescale considered here. Hence, we can neglect 
CO/H$_2$O kinetics in this model. The LiH molecule is a significant coolant only at $T\simlt 50$~K. More 
important influence on the thermal and chemical evolution is expected from dust and ionizing radiation. Dust 
particles confine a significant part of metals, produce additional cooling or heating, catalyze the H$_2$ 
formation as well. The ionizing radiation ionizes gas and dust. We are going to consider these effects elsewhere. 
But we note that a simple estimate gives that the H$_2$ formation on dust grains is inefficient within timescale 
considered here (see the next section). Several effects from the ionizing radiation on gas at $T>10^4$~K were 
described in \citep{v11}.


\section{Isochoric cooling rates}

\begin{figure}
\includegraphics[width=80mm]{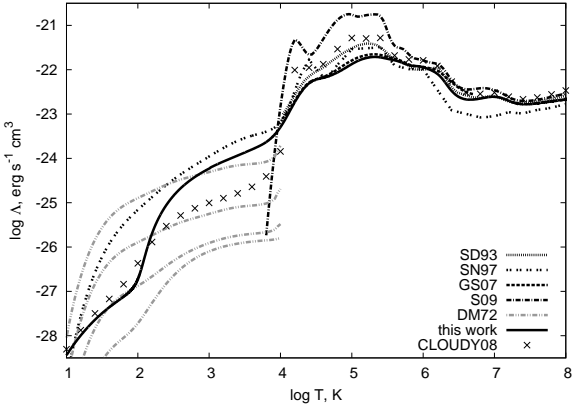}
\caption{
The cooling rates for solar metallicity. The isochoric rate { for gas with $n=1$~cm$^{-3}$}
calculated in this work is depicted by solid line. { The other lines correspond to the rates obtained in the previous calculations (see text for details).}
}
\label{figcoolsol}
\end{figure}

\begin{figure}
\includegraphics[width=80mm]{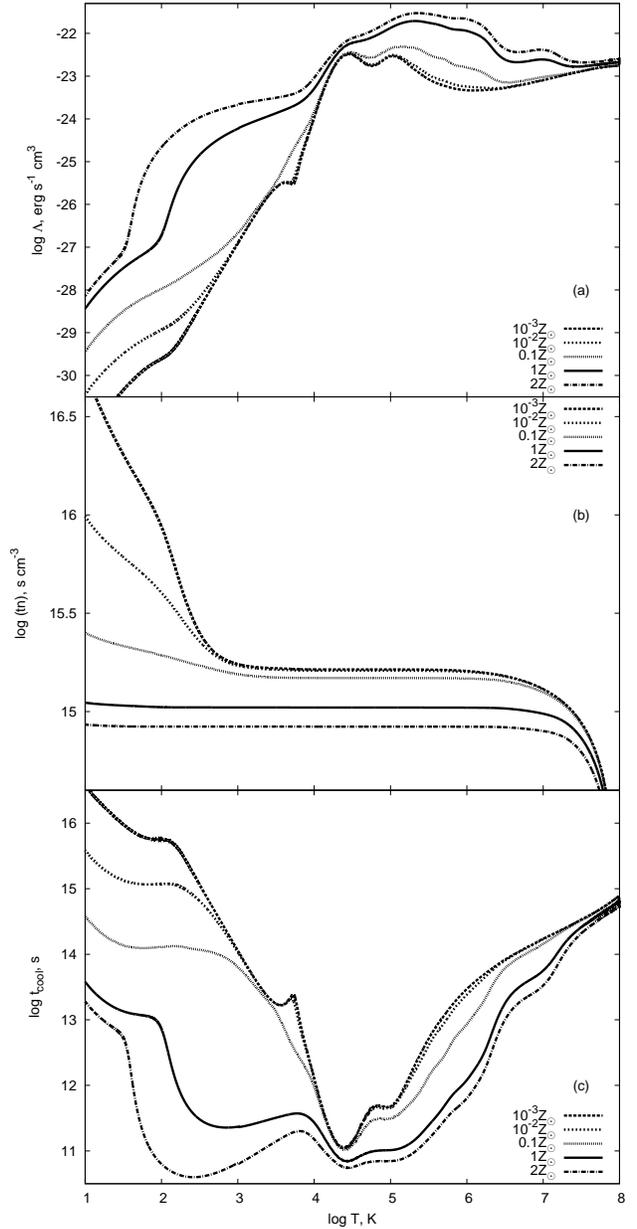}
\caption{
{
{\it Upper panel.} The isochoric cooling rates for gas with $10^{-4}$~$Z_\odot$, $10^{-3}$~$Z_\odot$,
$10^{-2}$~$Z_\odot$, $0.1$~$Z_\odot$, 1~$Z_\odot$, 2~$Z_\odot$ metallicities. The gas density is $n=1$~cm$^{-3}$.
{\it Middle panel.} The temperature dependence on the value of fluence, $\eta = \int n dt$ (for isochoric 
process it is simply $\eta = tn$), for the same gas parameters.
{\it Lower panel.} The isochoric cooling time.
}
}
\label{figcoolnet}
\end{figure}


For isochoric case we start our calculations from $T=10^8~K$ and follow it until the temperature becomes low as
$T=10~K$. { In general, the evolution of gas depends on initial ionic composition, temperature and density 
\citep[see e.g.][]{sd93,v12}, but here we will consider cooling of gas from very high temperature, so that there 
is no dependence on initial ionic composition of gas \citep[e.g.][ see the next section also]{v11}. }
Initially the ionization composition corresponds to the CIE at $T=10^8~K$ obtained from {\small CLOUDY}. 
Figure~\ref{figcoolsol} presents the isochoric cooling rates for a { gas with number density $n=1$~cm$^{-3}$} 
and solar metallicity. { The isochoric rate calculated in this work is depicted by solid line. To compare our
calculations of the rates to the previous ones we show the cooling rates obtained by \citet{sd93} (SD93, we have
choosen their non-equilibrium data), \citet{sn97} (SN97), \citet{gs07} (GS07), \citet{schure09} (S09), and
\citet{dalgarno72} for ionization fraction, $f_i=n_e/n_H = 10^{-4}$, $10^{-3}$, $10^{-2}$, $0.1$, which are 
shown by dash-dot-dotted lines from bottom to top. The data obtained from {\small CLOUDY} code is depicted by crosses
(the H$_2$ molecules are ignored in the equilibrium calculation).}
One should note that the cooling rates obtained by \citet{sd93}, \citet{sn97}, \citet{gs07} were got in non-equilibrium
calculations, so that these rates are more or less close to our rate within temperature range $10^4-10^6$~K.
\citet{sn97} didn't include high ionization states of metals in their model, so their cooling rate demonstrates
significant difference at $T\simgt 10^6$~K. The deviation at $T$ below $10^4$~K can be explained by different atomic 
data used in the calculations, and, as a consequence, difference of the ionization states of metals. The cooling 
curves obtained by \citet{schure09} and by using the {\small CLOUDY} code are examples of CIE rates, so that significant
difference of these rates from non-equilibrium calculations is not a surprise. The \citet{dalgarno72}
cooling curves were obtained by simply summing up cooling rates of hydrogen and metals, in which the fractional
ionization is used as a free parameter. One can note that our cooling rate is close to the \citet{dalgarno72} rate
for $x_e=0.1$ at $T\sim 10^3 - 10^4$~K and drops to that for $x_e = 10^{-4}$ at $T\simlt 10^2$~K. This proves a
necessity of self-consistent calculations of cooling rates. 

{ Figure~\ref{figcoolnet}(upper panel) presents non-equilibrium cooling rates for a dust-free collisionally controlled
gas in wide temperature ($10~K\le T\le 10^8~K$) and metallicity ($10^{-4}~Z_\odot \le Z \le 2~Z_\odot$) ranges. The
middle panel shows the temperature evolution (we present the dependence on fluence, $\eta = \int n dt$, which is for
isochoric process $\eta = tn$, where $t$ is time elapsed from the beginning of the evolution, $n$ is the number density
of gas). Here we should note that all collisional processes are proportional to the square of density. To see clearly
the importance of non-equilibrium effects the cooling time, $t_{cool} = k_b T/\Lambda(T)n$, is depicted at lower panel.
One can see that the cooling time is comparable with the time elapsed since the beginning of the evolution at very high
temperature, log~$T\simgt 7.5$, (early time of the evolution) for any metallicity and at low temperature,
log~$T\simlt 2.5$ for $Z \simlt 10^{-3}Z_\odot$. In other temperature and metallicity ranges the cooling time is shorter
than the evolution time, thus, the non-equilibrium effects are significant.}

\begin{figure}
\includegraphics[width=80mm]{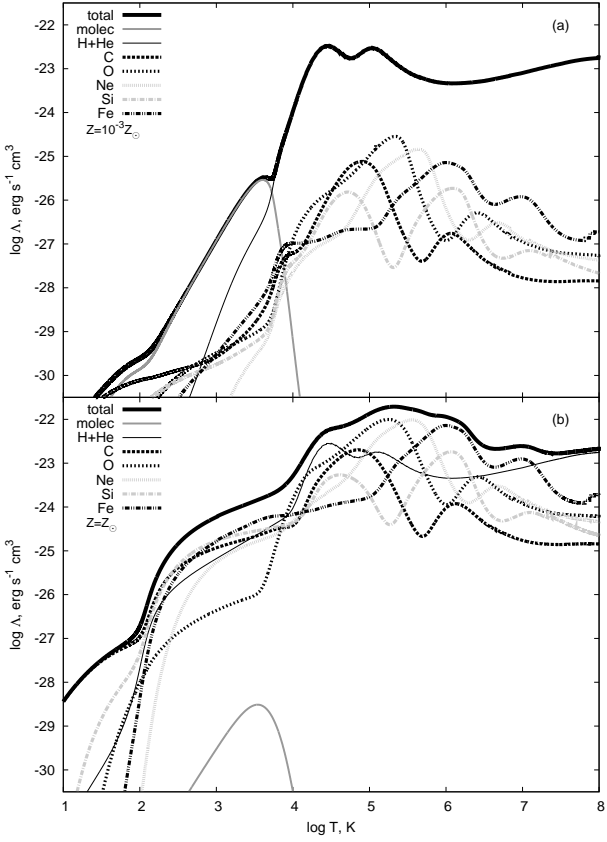}
\caption{
The contributions to the total isochoric cooling rate from each chemical element for $10^{-3}~Z_\odot$ 
(upper panel) and $Z_\odot$ (lower panel). The contribution by molecules (H$_2$ and HD) is depicted
by grey solid line, the total rate is presented by the thickest solid line, the deposit from H and He 
is shown by thin solid line.
}
\label{figcoold}
\end{figure}

\begin{figure}
\includegraphics[width=80mm]{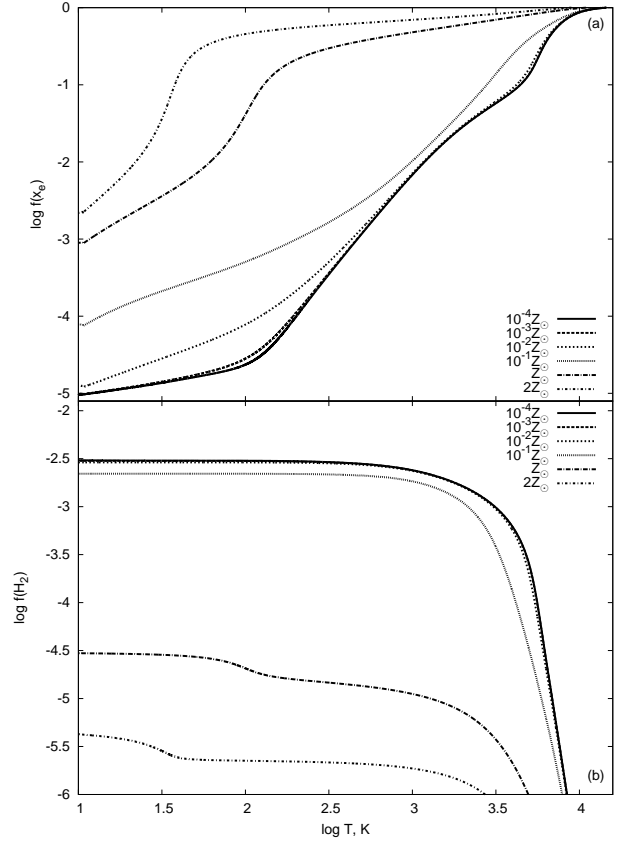}
\caption{
The electron, $x_e$, (upper panel) and molecular hydrogen, $x_{\rm H_2}$, (lower panel) fractions
for metallicities $10^{-4}$~$Z_\odot$, $10^{-3}$~$Z_\odot$, $10^{-2}$~$Z_\odot$, $0.1$~$Z_\odot$, 
1~$Z_\odot$ and 2~$Z_\odot$ in isochorically cooled gas.
}
\label{figxeh2}
\end{figure}

Figure~\ref{figcoold} shows the contributions to the total cooling rate from each chemical element and molecules 
for $10^{-3}~Z_\odot$ (upper panel) and $1Z_\odot$ (lower panel). Coolants at temperature higher than $10^4$~K are 
described by e.g. \citet{sd93,schure09,v11}. Also it is well-known that below $10^4$~K carbon, oxygen and H$_2$ molecules
dominate in cooling depending on metallicity of gas \citep[see, e.g.][]{hmk79,hmk89}. At very low metallicities, $Z\simlt10^{-3}~Z_\odot$, molecular hydrogen is the main coolant below $10^4$~K. For $Z\sim 10^{-3}~Z_\odot$ 
the contribution from carbon becomes significant only at $T \simlt 100$~K (upper panel). The increase of metallicity
leads to rise of contribution from metals and shortening of temperature range, where the H$_2$ dominates in 
cooling. For $Z\simgt 0.1~Z_\odot$ the H$_2$ deposit becomes minor at $T<10^4$~K, e.g. for solar metallicity 
the molecules produce a negligible cooling rate, whereas the contribution from metals is dominant (lower panel). 

One should note that thermal instability can develop at $T\simlt 10^4$~K. For $Z\simlt 10^{-2}~Z_\odot$ the instability
criterion for a gas cooled isochorically \citep{yua78,chevalier}
\be
f = {\rm d~ln~\Lambda \over d~ln~T} < 1 
\label{tiisoc}
\ee
is fulfilled around a small bump in the cooling rate at $T \sim (4-5)\times10^3$~K, which is produced due to the
contribution from molecular hydrogen (Figure~\ref{figcoold}a). The instability criterion is not reached within
$10^{-2}Z_\odot \simlt Z\simlt Z_\odot$. Again the conditions favoured to thermal instability can be found 
for $Z \simgt Z_\odot$ at $T \sim (0.8-7)\times10^3$~K (Figure~\ref{figcoold}b), where the instability
develops due to cooling by metals. 

The higher metallicity of a gas is the faster evolution goes, or in other words, the cooling time is decreased 
with the increase of metallicity, e.g. at $T\sim (1-10)\times 10^3$~K the cooling timescale for gas with solar
metallicity becomes shorter more than ten times compared to that scale for gas with $Z = 0.1~Z_\odot$ { (lower
panel of Figure~\ref{figcoolnet})}. Protons and electrons cannot recombine rapidly and such gas remains overionized: 
the electron fraction is high as 0.03 at 100K for solar metallicity (Figure~\ref{figxeh2}a). Such fast
evolution reveals in decrease of the H$_2$ fraction (although the H$^-$ fraction is high enough and depends on
metallicity weakly: it decreases from $\sim 1.4 \times 10^{-7}$ for $10^{-4}~Z_\odot$ to $5\times 10^{-8}$ for
$Z_\odot$, the H$_2^+$ fraction has the same order), so that for solar metallicity the H$_2$ fraction decreases 
more than 100 times compared to its 'universal' value, $(2-3)\times10^{-3}$, \citep{ohhaiman}, which is reached for 
$Z \simlt 0.1~Z_\odot$ (Figure~\ref{figxeh2}b). 

Here we consider the evolution starting from very high temperature, which initially corresponds to the conditions behind 
strong shock wave, $v\simgt 100$~km~s$^{-1}$. Usually it is assumed that dust grains existed in a gas before
such shock wave are destroyed behind it, { because the destruction time due to thermal sputtering. $\tau \simeq
2\times 10^4~{\rm yr}~(a/0.01\mu{\rm m})/n$ \citep[where $a$ is the size of the dust particle, see][]{dustde}, is shorter
than the cooling time for a gas cooled from $T\simgt 10^6$~K depicted in Figure~\ref{figcoolnet}c. Moreover, the
accretion growth rate is negligible compared to sputtering rate \citep[see Figure 7 in][]{dustde}, so that dust
grains have no time to reform before the final time is reached in our calculations.} 
Hence, in the present calculations we can do not take into account the H$_2$ formation on dust grains. 

But if dust grains nevertheless exist in a gas, then it is well-established that they catalyze the H$_2$ 
production in the interstellar medium \citep{hmk79}. In the warm neutral gas ($T\simgt 10^3$~K) the H$_2$ 
formation of grains dominates over that due to the gas-phase reactions at $Z\simgt 0.1~Z_\odot$ \citep[see 
the analysis in][]{gloverh2dust}. The time of H$_2$ formation on grain-surface is $t_{grains}^{\rm H_2}
\sim 3 \times 10^{16} (T/100)^{-1/2}\, (n Z/Z_{\odot})^{-1}$s \citep{hmk79}\footnote{{ A recent analysis of H$_2$
formation on grain surfaces by \citet{cazaux04} that takes both physisorbed and chemisorbed hydrogen into account
demonstrates that in the conditions of interest in this paper the computed H$_2$ formation rate is very similar 
to the widely used rate of \citet{hmk79}}}. The H$_2$ formation process can be effective, if the formation 
time is shorter than { the time elapsed from the beginning of the evolution presented in Figure~\ref{figcoolnet}b. 
More exactly, we are interested in the part of the evolution starting from the moment, when the temperature reaches
$T\simeq 10^4$~K, to the end the calculation, when $T=10$~K. Just in this temperature range hydrogen recombines
efficiently and H$_2$ molecules can form on grain surface. This timescale is the same order of magnitude as the
full time of the calculation (Figure~\ref{figcoolnet}b). One can see that the $t_{grains}^{\rm H_2}$ is about 1-2 orders
of magnitude longer than the time needed to cool from $T\sim 10^4$~K to $10$~K. So that, we note that neither
grain-surface nor gas-phase H$_2$ formation channel is effective in a gas with $Z> 0.1~Z_\odot$, and the H$_2$ fraction
remains negligible (see lower panel of Figure~\ref{figxeh2}). }

In general, if dust grains survive after strong shock waves, they contain a significant part of metals. 
In this case the evolution time of a gas becomes longer and, as a consequence, grain-surface molecule
formation becomes more efficient. Because of several additional free parameters, e.g. dust destruction efficiency,
depletion factor, we will study the effects from dust on thermal evolution elsewhere. 

Another molecules, which can significantly affect the thermal evolution of gas, are CO and H$_2$O \citep[e.g.,][]
{hmk79,glover07}. { These molecules mainly form via reaction channels initiated by the hydroxyl OH \citep[e.g.,][]
{hmk79}. The OH radical forms through reaction between neutral oxygen and molecular hydrogen. Then, the lack
of H$_2$ molecules produces smaller OH fraction. Moreover, as it is mentioned above} the H$_2$ formation on dust grains is
inefficient in the conditions considered here, and its fraction decreases with metallicity. The CO formation timescale 
(for H$_2$O molecule the timescale is close) is $t({\rm O\rightarrow CO}) \simeq (10^{-3}/x_{\rm
H_2}) (100{\rm cm^{-3}}/n) (Z_\odot/Z)$~Myr \citep{glover07}. Comparing $t({\rm O\rightarrow CO})$ with the cooling
time, which is { $t_{cool} \sim 0.01~{\rm Myr}~(1{\rm cm^{-3}}/n)$ at $T\sim 10^3-10^4$~K for solar metallicity
(Figure~\ref{figcoolnet}c)}, it is easy to see that the CO and H$_2$O molecules will not form in quantities large
enough to dominate the cooling. 

The higher gas density is the smaller recombination timescales of hydrogen and metals become, so the above-described 
picture is expected to change in isobaric process.


\section{Isobaric cooling rates}

First of all we should choose initial temperature and density of a gas for isobaric calculations.
{
Certainly, to compare with the isochoric models we should start our calculations from $T=10^8$~K. However 
we extend our calculations to $T<10^4$~K, in this range main radiative losses are through fine-structure lines 
of metals. The level populations of these lines go from equilibrium to non-equilibrium within $n\sim 1-100$~cm$^{-3}$ 
at $T\simlt 10^3$~K. To catch this transition we should set initial density $n\simlt 10^{-5}$~cm$^{-3}$ for isobaric
calculations started from $T=10^8$~K. Such density, which is relevant for the intergalactic medium, is quite 
low for the interstellar medium and, moreover, the cooling time of such gas is larger than the current age 
of the Universe. For higher density isobaric cooling rates calculated from $T=10^8$~K have negligible 
dependence on number density. Hence, to study the transition of level populations of metal fine-structure 
lines from non-equilibrium to equlibrium we need to take lower initial temperature. But this temperature 
should be high enough, that initial difference between isochoric and isobaric cooling rates is small 
and dependence of the evolution on initial temperature is negligible. Using these conditions we can 
constrain the gas temperature from several$\times 10^5$ to $10^6$~K \citep{gs07,v12}. In this case the 
compression factor (the ratio of initial temperature to $10^3$~K) is about $\simlt 10^3$ and we can 
consider number densities $n \simgt 10^{-3}$~cm$^{-3}$ as initial values.
}

\begin{figure}
\includegraphics[width=80mm]{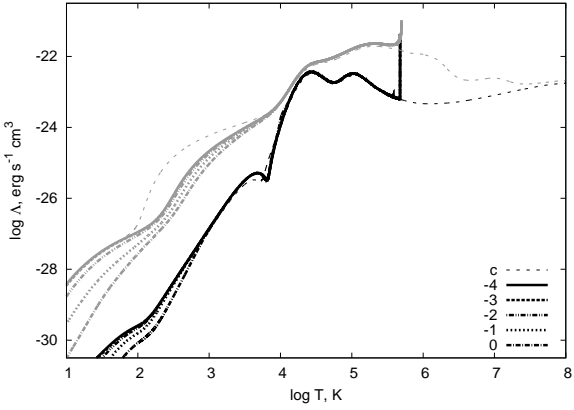}
\caption{
The isobaric (thick lines, numeric labels) and isochoric (thin lines, lablel 'c') cooling rates for metallicities
$10^{-3}$~$Z_\odot$ (black lines), 1~$Z_\odot$ (grey lines). The numeric labels are the logarithm of the initial
number density. The initial temperature for the isobaric calculations equals $5\times 10^5$~K.
{ The density of isochorically cooled gas is $0.1$~cm$^{-3}$.}}
\label{figcoolb}
\end{figure}

{ 
Also we should mention that we consider isobaric compression, which is natural for a gas parcel crossed
through strong shock front. So the gas temperature ahead a shock is much lower than the value $T = 3 m_p
v^2/16 k$, where $v$ is a shock front velocity. To mimic this we assume that the ionic composition
in a gas before a shock corresponds to that at $T = 2\times 10^4$~K.
}
Since here we consider pure collisional ionization model, we should constrain shock wave velocity by a 
value that does not lead to the radiative precursor formation. A stable photoionization precursor will be 
formed for shock velocity higher than $v_*\simgt 175$~km~s$^{-1}$ \citep{sd96} or shock temperature $T_* 
= 3 m_p v_*^2/16 k \simgt 7\times 10^5$~K. We have found that the cooling rate of a gas behind a shock 
wave with $T_s \simeq 5\times 10^5$~K tends rapidly to the rate in a gas collisionally cooled from $T=10^8$~K
\citep{v12}. Thus, { taking into account the above-mentioned constrains on gas temperature} we can choose 
$T=5\times 10^5$~K as initial temperature. 

In general, the transition from isobaric to isochoric cooling is controlled by dynamical time of a system: 
cooling is isobaric when the cooling time is much greater then the dynamical time of a system, $t_c/t_d \gg 1$ 
and it is isochoric when the time ratio is opposite, $t_c/t_d \ll 1$. For a example, the evolution of a
density fluctuation (e.g., a cloud) is determined by the transition from isobaric to isochoric cooling 
\citep[e.g.,][]{gs07}. Such transition is able to determine the characteristic size and mass of the cold 
dense clumps in two-phase medium \citep{burkertclouds}. 

Figure~\ref{figcoolb} presents the isobaric cooling rates for gas with metallicity $10^{-3}~Z_\odot$
and $1~Z_\odot$ (thick lines) and the initial number densities $n = 10^{-4}, 10^{-3}, 10^{-2}, 0.1$ 
and 1~cm$^{-3}$ (the numeric labels in the Figure are the logarithm of the initial density). As 
it is mentioned above the initial temperature for the isobaric calculations equals $5\times 10^5$~K. 
{ 
First of all, the tails at $T\sim 5\times 10^5$~K are connected with the relaxation of gas behind 
a shock front \citep{v12}. During this relaxation gas is rapidly ionized, the ionic composition corresponded
initially to $T=2\times 10^4$~K tends rapidly to that at $T\sim 5\times 10^5$~K. The duration of the period
corresponded to almost constant temperature is very short $\sim 0.01t_f$, where $t_f$ is the total evolution 
time, which is elapsed from the beginning to that at which the temperature reaches $10$~K. But more or less
full relaxation reaches during $t\sim 0.1 t_f$. 
}

\begin{figure}
\includegraphics[width=80mm]{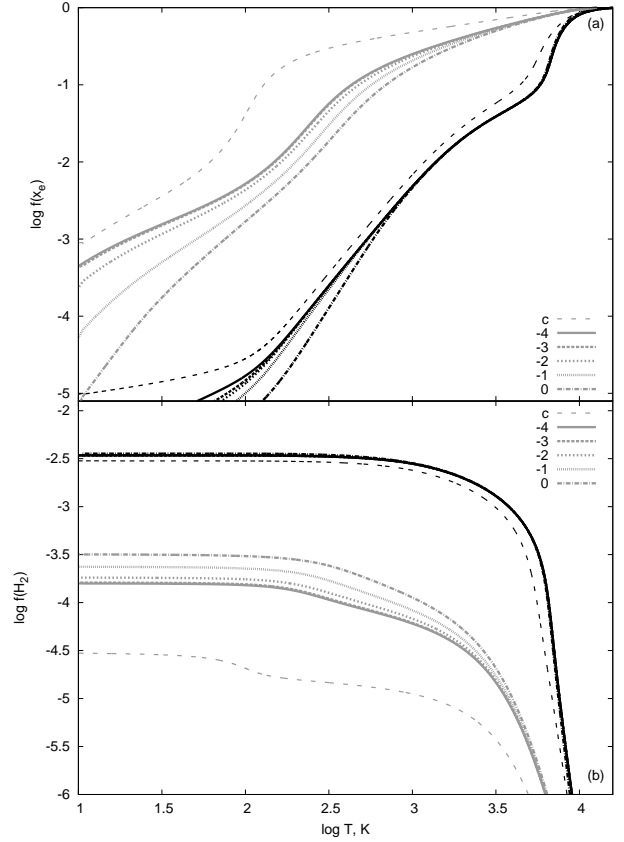}
\caption{
The electron $x_e$ (upper panel) and molecular hydrogen $x_{\rm H_2}$ (lower panel) fractions
for metallicities $10^{-3}$~$Z_\odot$ (black lines), 1~$Z_\odot$ (grey lines) in isobarically 
cooling gas. The numeric labels are the logarithm of the initial number density. The initial 
temperature equals $5\times 10^5$~K. { The label 'c' corresponds to the isochorically cooled
gas with $n=0.1$~cm$^{-3}$.}
}
\label{figxeh2b}
\end{figure}

In Figure~\ref{figcoolb} one can see that the difference between isochoric and isobaric cooling rates reveals 
at $T\simlt 10^4$~K. Such difference can be explained by two reasons. At first, the recombination of hydrogen and 
metals becomes more efficient in a gas with higher density. At second, the cooling below 
$10^4$~K is mainly due to molecule roto-vibrational and metal fine-structure transitions, whose level populations 
go from non-equilibrium ($L\sim n^2$) to the equilibrium ($L\sim n$) in gas with $n\sim 1-10^4$~cm$^{-3}$. 
The first (the hydrogen recombination) is significant at $T\sim 10^3 - 10^4$~K, the second (the equilibration) plays 
a role at $T\simlt 400$~K. 

One can see that for isobarically evolved gas the thermal instability can develop at $T\simlt 10^4$~K. The
instability criterion is weaker than that for isochorically cooling gas \citep{yua78,chevalier}
\be
f = {\rm d~ln~\Lambda \over d~ln~T} < 2.
\label{tiisob}
\ee
For $Z\simlt 10^{-2}~Z_\odot$ this criterion is fulfilled around a small bump in the cooling rate at $T \sim
(4-5)\times10^3$~K, which is produced due to the contribution from molecular hydrogen. For gas with $Z \simgt 
Z_\odot$ and initial density $n=10^{-4}$~cm$^{-3}$ at $T=5\times 10^5$~K the temperature range, where the 
thermal instability can develop, is $T \sim (0.2-10)\times10^3$~K. This range becomes narrower with increase 
of initial density, which is explained by steeping of cooling curve (Figure~\ref{figcoold}b) due to both more 
efficient hydrogen recombination and equilibration of the fine-structure level populations.

\begin{figure}
\includegraphics[width=80mm]{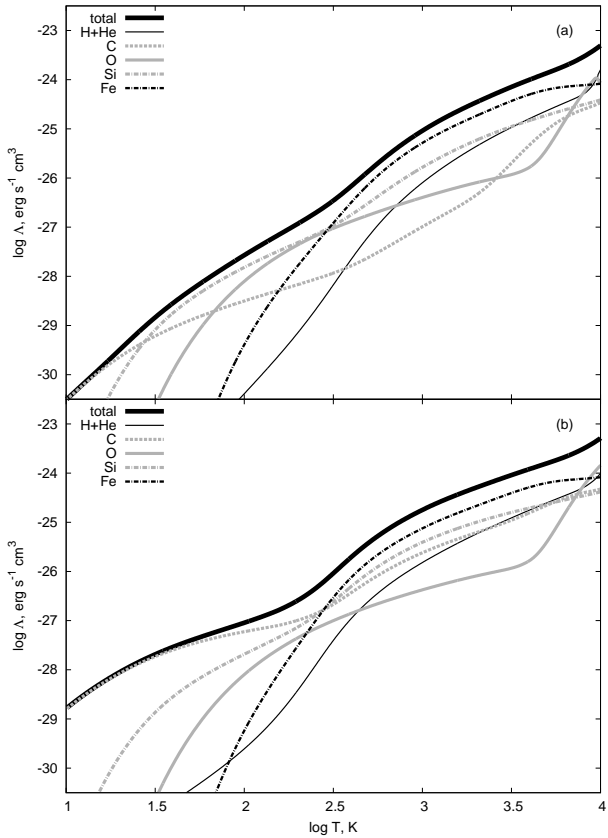}
\caption{
The contributions to the total isobaric cooling rate from major chemical elements for initial number 
density $1$~cm$^{-3}$ (upper panel) and $10^{-2}$~cm$^{-3}$ (lower panel) for a gas with $Z=Z_\odot$. 
The initial temperature equals $5\times 10^5$~K.}
\label{figcooldb}
\end{figure}

Figure~\ref{figxeh2b} shows the electron (upper panel) and molecular hydrogen (lower panel) fractions for metallicities
$10^{-3}$~$Z_\odot$ (black lines), 1~$Z_\odot$ (grey lines) in gas cooled isobarically. The numeric labels are the
logarithm of the initial density. The increase of density intensifies both hydrogen recombination and H$_2$ molecule formation.  For solar metallicity gas the decrease of ionization fraction compared to that for gas cooled isochorically
is significant and reaches 0.5-2 order of magnitude at $T\simlt 10^3$~K (compare the thin dash line to the thick ones in
the upper panel). So that the higher initial density is the larger deviation from the isochoric case can be found. The H$_2$ fraction at $Z=Z_\odot$ increases about 10 times (thick grey lines in the lower panel), but the H$_2$ contribution
to the cooling rate still remains negligible for solar metallicity. We should note that although the H$_2$ fraction is 
higher than that in the isochoric case, but this is still insufficient for CO/H$_2$O formation. For $10^{-3}~Z_\odot$ 
the dependence of ionization kinetics on initial gas density is less pronounced (thick black lines in the upper panel). 
It can be appreciable only at $T\simlt 100$~K. The H$_2$ fraction reaches its 'universal value' and does not depend on
density (thick black lines in the lower panel).

\begin{figure}
\includegraphics[width=80mm]{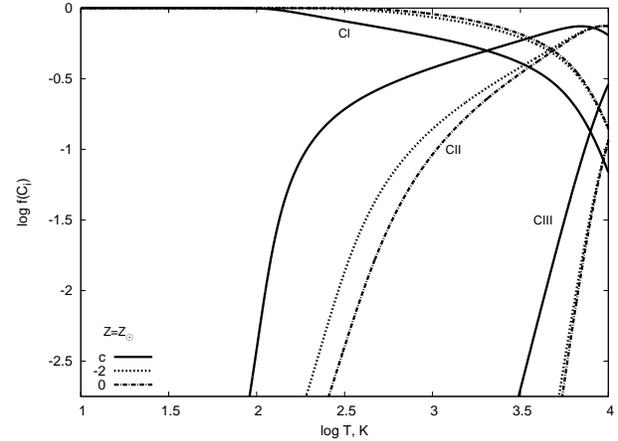}
\caption{
{
The CI and CII fractions in isochorically (solid line) and isobarically (dot and dash-dotted lines) 
cooled gas with solar metallicity. In the isochoric calculation the density of a gas is 0.1~cm$^{-3}$ (the 
pressure equals $1.38\times 10^{-13}(T/10^4{\rm K})$~erg~cm$^{-3}$). In the isobaric models the 
initial density values are $10^{-2}$~cm$^{-3}$ (the pressure is $6.9\times 10^{-13}$~erg~cm$^{-3}$)
and $1$~cm$^{-3}$ (the pressure is $6.9\times 10^{-11}$~erg~cm$^{-3}$), depicted by dot and dash-dotted 
lines, respectively. The initial temperature for the isobaric calculations equals $5\times 10^5$~K.
}
}
\label{figcarb}
\end{figure}

Figure~\ref{figcooldb} shows the contributions to the total isobaric cooling rate from major chemical elements 
for initial number densities $n=1$~cm$^{-3}$ (upper panel) and $n=10^{-2}$~cm$^{-3}$ (lower panel) for gas with 
$Z=Z_\odot$ (the contributions by N, Ne, Mg and molecules are minor in this temperature range for gas with 
such metallicity). We present the low-temperature part of the cooling curve, because the isobaric effects 
reveal in this range. One can note a strong difference between cooling rates at $T\simlt 300$~K.
The contribution by carbon, which is the main coolant for the initial density $10^{-2}$~cm$^{-3}$, is 
significantly suppressed for higher initial density value. One can find that at $T\simlt 300$~K the energy 
losses in the fine-structure transition of CI (609~$\mu$m) dominate in the cooling, and also that the critical 
density (when the level populations go from non-equilibrium, $L\sim n^2$, to equilibrium, $L\sim n$) of this 
transition is $160(T/100)^{-0.34}$~cm$^{-3}$ \citep{hmk89}. During isobaric contraction from $5\times 10^5$~K 
to $\simlt 300$~K the density increases more than $\simgt 2.3\times 10^3$ times. So that at $T\simlt 300$~K the 
cooling of gas with initial density $n = 10^{-2}$~cm$^{-3}$ should be more efficient (where $L\sim n^2$) than that of
gas with initial density $n = 1$~cm$^{-3}$ (where $L\sim n$). 

As an example we consider the carbon ionization kinetics in detail. { Figure~\ref{figcarb} present the carbon 
ionization states for a gas with solar metallicity cooled isochorically (solid line) and isobarically (dot 
and dash-dotted lines). 
In the isobaric models the CI fraction is about 1 at $T\simlt 300$~K. Then, there are no other coolants 
except CI in this temperature range (Figure~\ref{figcooldb}), and the difference between the cooling rates 
(Figure~\ref{figcooldb}) is explained by the equilibration of the CI level populations. Similar effect can 
be found for CII within $T\sim 300 - 10^4$~K, but it is less pronounced 
than that for CI. In general, a decrease of pressure (or initial density with fixed initial temperature) 
leads to that the isobaric cooling rate tend to the isochoric one (see e.g. Figure~\ref{figcoolb}). For
low-metallicity gas the role of metals to cooling rate diminishes and the transition becomes less distinguishable,
because the critical density for the H$_2$ roto-vibrational levels is higher, it is about $10^4$~cm$^{-3}$. In the 
presence of UV radiation this transition can be significant in diffuse clouds in the interstellar medium 
\citep[see e.g.][]{carbonism}.
}

\section{ { Application: supernova explosion} }

In the previous sections we have considered the cooling rates of gas cooled from $T\simgt 5\times 10^5$~K. These 
conditions naturally correspond to a gas behind strong shock waves originated in collisions of galaxies, virial 
shocks in clusters of galaxies and, more common and usual process in the interstellar gas, supernova (SN) explosions.
The temperature of gas just behind SN shockwave is generally higher than $5\times 10^5$~K, so that the cooling rates
calculated in this paper are valid for physical conditions behind SN shockwave. To study the SN evolution we use the
multi-dimensional parallel code ZEUS-MP \citep{zeusmp}. We have added the cooling processes into the energy equation
using the explicit four-step Runge-Kutta method. 

The upper panel of Figure~\ref{figcoolsol} shows that in the high-temperature regime ($T\simgt 10^4$~K) the cooling 
rate calculated in this work is close to that obtained in other studies. A significant difference of the rates can 
be found in the low-temperature regime ($T\simlt 10^4$~K). Certainly, the difference between cooling rates more than 
2-3 times can lead to earlier (or later) beginning of the radiative phase of a SN shell and, as a result, different 
scales of fragments appeared due to thermal instability. 
In general, gasdynamic simulations with self-consistent chemical kinetics undoubtedly trace any variation of cooling 
rate due to both chemical transformation and gas dynamics, but such calculations are very time-consuming, restricted 
in resolution (one should include multi-species advection in gas dynamics) and in applications (one should take into
account valid chemical species and processes for a given physical task). 
 
Use of tabulated cooling rates gives more flexibility in study of some physical processes. One of them is mixing
metals in the interstellar/intergalactic medium \citep{mix00,avillez02,mix04,scalo04mix,mix09,burkertmix,mix12}. 
Metals are produced in stellar interiors and ejected into surroundings by supernova explosions. 
But available data for cooling rates \citep{sd93,gs07,schure09,wiersma} are very constrained in metallicity and
temperature, or even combined from different sources \citep{schure09}. Of course, one can utilize pre-computed tables
assumed the equilibrium \citep[e.g. use {\small CLOUDY}, see][]{wiersma} or choose several species for nonequilibrium
calculations, whereas others are considered in equilibrium \citep[e.g.][]{smithTI}. But in the former we have 
equlibrium rates, which significantly differ from the nonequilibrium ones, whereas the latter needs a significant computational time.

\begin{figure*}
\includegraphics[width=160mm]{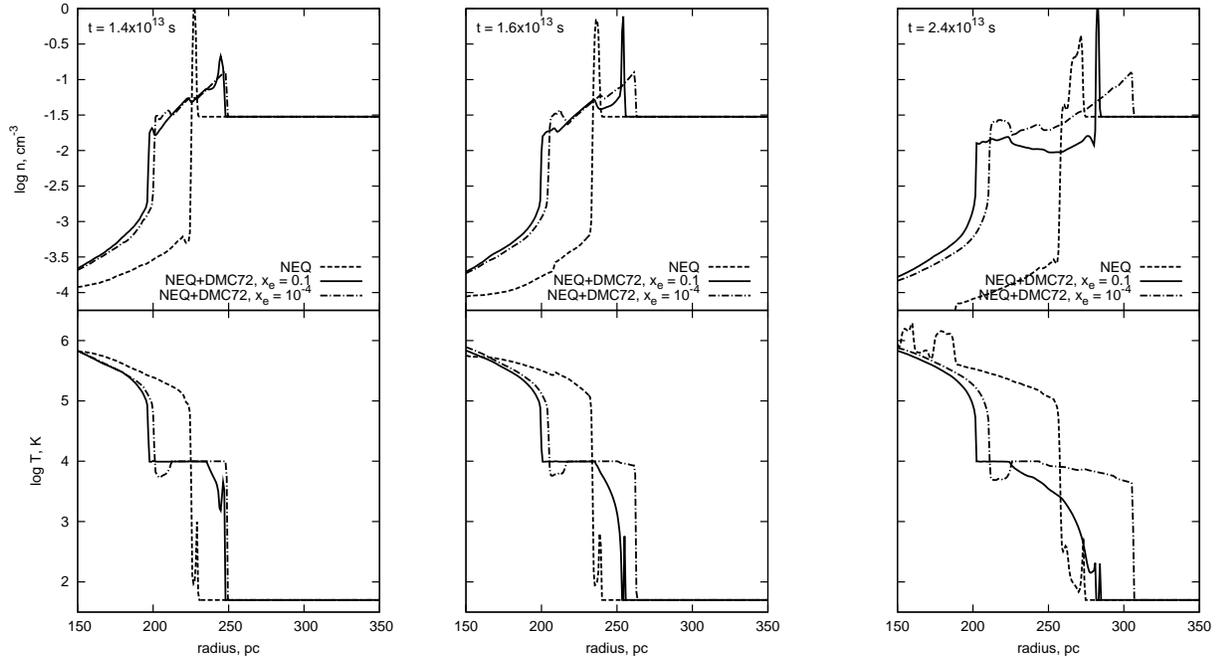}
\caption{
{ 
The radial distributions of density (upper panels), temperature (lower panels) around a SN exploded 
in a homogeneous medium with $n=3\times10^{-2}$~cm$^{-3}$ and solar metallicity for cooling rates in low 
temperature range, $T<10^4$~K: 
a) the self-consistent calculations presented in this paper (solid lines), b) the \citet{dalgarno72} rate for 
the fixed electron fraction $x_e = 0.1$ (dashed lines) and c) the same rate, but for $x_e = 10^{-4}$ (dotted 
lines). Left column panels shows the distributions at $t=1.4\times 10^{13}$~s after explosion, middle column
presents the same at $t=1.6\times 10^{13}$~s and right column of panels shows the same at $t=2.4\times 10^{13}$~s. 
In high temperature range, $T>10^4$~K, we use the cooling rates calculated in this work.
}
}
\label{figsn}
\end{figure*}

Besides our cooling rates Figure~\ref{figcoolsol} presents two other cooling curves at $T<10^4$~K: \citet{sn97}
and \citet{dalgarno72}. The former differs slightly from the rate calculated here at $T\simgt 10^2$~K, so we
don't consider it. The latter widely used in the ISM studies \citep[e.g.,][]{avillez02,burkertmix} is valid only 
at $T<10^4$~K and demonstrates significant differences from ours for $x_e \simlt 0.1$. So that we use the cooling
rates calculated in this paper in high temperature range, $T>10^4$~K, whereas for $T<10^4$~K we take three options 
for cooling rate: a) our self-consistent isochoric calculations (label 'NEQ'),  b) the \citet{dalgarno72} rate for 
the fixed electron fraction $x_e = 0.1$ (label 'NEQ+DMC72, $x_e=0.1$') and c) the same rate, but for $x_e = 10^{-4}$ 
(label 'NEQ+DMC72, $x_e=10^{-4}$'). Our choice of $x_e$ is explained that the rate for higher value $x_e$ is close to 
our rate at $T\sim 10^3-10^4$~K, whereas the lower $x_e$ is close to the value that \citet{smithTI} found for the
interstellar gas with solar abundance \citep[the same value can be obtained for gas ionized by cosmic rays in the 
local ISM, see e.g.][]{wolfire95}.

Here we investigate the dependence of the SN shell evolution in a homogeneous medium with solar metallicity on
different cooling rates in low temperature range, $T<10^4$~K. 
We have carried out the 2D gasdynamic simulations (cylindrical geometry) of a SN explosion. The numerical resolution
is $1000\times 1000$, that corresponds to the physical size of a cell 0.4~pc. The SN energy is initially injected in
the central region with radial size 3.2~pc, the total energy is $10^{51}$~erg. To exclude the evolution of the
gas in the medium before shockwave we set the temperature of this gas to 50~K and do not allow to cool it down. The
differences of the SN shell evolution due to variation of the cooling rate can be found just after time $t\sim
10^{13}$~s, when initially hot shell cooled to $\sim 10^4$~K. We continued the calculations as long as $t=10^{14}$~s,
at this time various instabilities destroy SN shell, but here we are interested in the beginning of the shell
evolution cooled below $10^4$~K. At $t\simlt 3\times 10^{13}$~s the shell is still close to spherical shape, so 
that we can present radial distributions of physical quantities only.

Figure~\ref{figsn} present the radial distributions of density (upper) and temperature (lower) around a SN exploded 
in a homogeneous medium with density $n=3\times10^{-2}$~cm$^{-3}$ and solar metallicity for different cooling rates 
at $T<10^4$~K. At $t=1.4\times 10^{13}$~s (the left row of panels in Figure~\ref{figsn}) one can find that the 
radiative phase begins in the SN shell cooled with the rate calculated in this work ('NEQ') and the \citet{dalgarno72} 
rate for the fixed electron fraction $x_e = 0.1$  ('NEQ+DMC72, $x_e=0.1$'). In the first ('NEQ') model the thin shell 
is formed, the density in the shell increases in several times and the temperature is low as $\sim 100$~K. Whereas 
in the second ('NEQ+DMC72, $x_e=0.1$') model the effects of radiative cooling begin to become apparent only: the
radiative phase have just begun. The main difference between two models is the absence of temperature plateau at
$T\simeq 10^4$~K in case of the self-consistent cooling rate ('NEQ'). The plateau (see radial distances $r\sim
200-250$~pc) originates from lower cooling rate in the 'NEQ+DMC72, $x_e=0.1$' model in comparison with the rate used
in the 'NEQ' model. At $t=1.6\times 10^{13}$~s the SN shell becomes radiative for the 'NEQ+DMC72, $x_e=0.1$' model. 
A gas in the shell begins to cool below $10^4$~K, and the plateau in the radial temperature distribution disappears gradually. In the third ('NEQ+DMC72, $x_e=10^{-4}$') model considered here a gas in the shell starts to cool down 
below $10^4$~K only at $t=2.4\times 10^{13}$~s (right column of panels in Figure~\ref{figsn}). This delay can be easily
explained by lack of cooling at $T\sim 10^4$~K in the 'NEQ+DMC72, $x_e=10^{-4}$' model in comparison with the other
models considered in this section (Figure~\ref{figcoolsol}). Note that this delay can significantly affect on the SN 
hot bubble dynamics: the hot bubble has different sizes depending on the cooling curve assumed in calculation. This 
can be apparent in mixing metals, because the metals ejected by SN are initially confined in the hot bubble. Thus, we
can conclude that tabulated cooling rates should be used carefully, because some gaseous structures may form due to
using cooling rates cooling rates calculated for different physical conditions. As far as possible one should use cooling rates calculated self-consistently for the whole temprature range of one's interest.


\section{Conclusions}

\noindent

In this paper we have presented self-consistent calculations of the non-equilibrium cooling rates of a dust-free
collisionally controlled gas with metallicities in the range $10^{-4}~Z_\odot \le Z \le 2~Z_\odot$ cooling from 
high temperatures $T>5\times10^5$~K down to $10$~K\footnote{The isochoric cooling rates presented in
Figure~\ref{figcoolnet}b are available in electronic form at http://ism.rsu.ru/cool/coolfun.html.}. 
We have found that
\begin{itemize}
 \item molecular hydrogen dominates cooling at $10^2 \simlt T\simlt 10^4$~K and $Z\simlt 10^{-3}~Z_\odot$, its
 contribution around $T\sim (4-5)\times 10^3$~K stimulates thermal instability at $Z\simlt 10^{-2}~Z_\odot$;
 \item the ionization fraction remains rather high at $T\simlt10^4$~K: it reaches up to $\sim 0.01$ at $T\simeq 
 10^3$~K and $Z\simlt 0.1~Z_\odot$, and becomes higher for higher $Z$;
 \item despite high ionization fraction the abundance of molecular hydrogen decreases with metallicity;
 \item at $T\simlt 10^4$~K isobaric cooling rates are lower than isochoric ones;
 \item in isobaric processes the ionization fraction decreases considerably and molecular hydrogen grows more 
 efficient compare to isochoric ones.
\end{itemize}
We have compared our self-consistent cooling functions with those published in previous works and found considerable
differences. In particular, we found that the two approaches for the cooling processes result in a fairly distinct
dynamics of supernova remnants.

\section{Acknowledgements}

\noindent
{ The author thanks the anonymous referee for valuable comments. } 
The author is grateful to Yuri Shchekinov for help and many useful discussions,
Mikhail Eremin and Eduard Vorobyov for stimulating discussions, Ilya Khrykin for
his assistance. 
Gary Ferland and {\small CLOUDY} community are acknowledged for creating of the excellent tool
for study photoionized plasma.
This work is supported by the RFBR through the grants 12-02-00365, 12-02-00917, 
12-02-90800, and partially 12-02-92704, 11-02-01332, 11-02-97124, and by the Russian 
federal task program "Research and operations on priority directions of development of 
the science and technology complex of Russia for 2009-2013" (state contract 14.A18.21.1304).
The author is grateful for support from the "Dynasty" foundation.


\appendix

\section[]{Molecular hydrogen and deuterium chemistry}

{ 
In this section we present the minimal model of chemical kinetics for molecular hydrogen and deuterium in the 
low temperature range, $T<2\times 10^4$K. We include collisional gas-phase reactions only. The reactions are 
listed in Table~\ref{table2}. This model is based on the well-tested minimum models for molecular hydrogen 
and deuterium chemistry for both primordial \citep{abel97,galli98} and metal-enriched \citep{omukai05,glover07} 
gas. We don't list here the helium chemistry, because it consists of collisional ionization and recombination
only, these processes are inevitable part of our model and the references to reactions can be found in \citep{v11}.
Other helium species, like HeH$^+$, cannot change thermal evolution of a gas significantly, so that we
do not include its in the model. } 

\begin{table}
\caption{Chemical reaction rates}
\center
\begin{tabular}{lc}
\hline
\hline
reaction              & Reference \\
\hline
H$_2$ chemisrty & \\
\hline
${\rm H     +   e^-  \to H^-   +   h\nu}$    &\citet{galli98}    \\  
${\rm H^-   +   h\nu \to H     +   e^-}$     &--                 \\  
${\rm H^-   +   H    \to H_2   +   e^- }$    &--                 \\  
${\rm H^-   +   H^+  \to H_2^+ +   e^- }$    &--                 \\  
${\rm H^-   +   H^+  \to 2H            }$    &--                 \\  
${\rm H^+   +   H    \to H_2^+ +   h\nu}$    &--                 \\  
${\rm H_2^+ +   H    \to H_2   +   H^+ }$    &--                 \\  
${\rm H_2^+ +   e^-  \to 2H            }$    &--                 \\  
${\rm H_2   +   H^+  \to H_2^+ +   H   }$    &--                 \\  
${\rm H_2   +   e^-  \to 2H    +   e^- }$    &--                 \\  
${\rm H_2   +   H    \to 3H            }$    &\citet{maclow86}   \\  
${\rm H^+   +   e^-  \to H     +   h\nu}$    &\citet{verner96}   \\  
${\rm H     +   e^-  \to H^+   +   2e^-}$    &\citet{abel97}     \\  
${\rm H^-   +   e^-  \to H     +   2e^-}$    &--                 \\  
${\rm H^-   +   H    \to 2H    +   e^- }$    &\citet{sk87}       \\  
\hline
D chemistry & \\
\hline
${\rm D^+   +   e^-  \to D     +   h\nu}$    &\citet{galli98}    \\  
${\rm D     +   h\nu \to D^+   +   e^- }$    &--                 \\  
${\rm D     +   H^+  \to D^+   +   H   }$    &--                 \\  
${\rm D^+   +   H    \to D     +   H^+ }$    &--                 \\  
${\rm D     +   H_2  \to H     +   HD  }$    &--                 \\  
${\rm D^+   +   H_2  \to HD    +   H^+ }$    &--                 \\  
${\rm HD    +   H    \to H_2   +   D   }$    &--                 \\  
${\rm H^+   +   HD   \to H_2   +   D^+ }$    &--                 \\  
${\rm D     +   e^-  \to D^-   +   h\nu}$    &--                 \\  
${\rm D^+   +   D^-  \to 2D            }$    &--                 \\  
${\rm H^+   +   D^-  \to D     +   H   }$    &--                 \\  
${\rm H^-   +   D    \to H     +   D^- }$    &--                 \\  
${\rm D^-   +   H    \to D     +   H^- }$    &--                 \\  
${\rm D^-   +   H    \to HD    +   e^- }$    &--                 \\  
\hline
\hline
\end{tabular}%
\label{table2}
\end{table}

\end{document}